\documentstyle[12pt]{article}

\newcommand{\sect}[1]{\setcounter{equation}{0}\section{#1}}

\def\be{\begin{equation}}
\def\ee{\end{equation}}
\def\ba{\begin{eqnarray}\samepage}
\def\ea{\end{eqnarray}}

\font\twelvemsa=msam10 scaled 1200

\font\sevenmsa=msam7
\font\fivemsa=msam5
\newfam\msafam
\textfont\msafam=\twelvemsa
\scriptfont\msafam=\sevenmsa
\scriptscriptfont\msafam=\fivemsa
\def\msa{\ifcase\msafam 0\or1\or2\or3\or4\or5\or6\or7\or8\or9\or A\or B\or
C\or D\or E\or F\fi}

\font\twelvemsb=msbm10 scaled 1200

\font\sevenmsb=msbm7
\font\fivemsb=msbm5
\newfam\msbfam
\textfont\msbfam=\twelvemsb
\scriptfont\msbfam=\sevenmsb
\scriptscriptfont\msbfam=\fivemsb
\def\msb{\ifcase\msbfam 0\or1\or2\or3\or4\or5\or6\or7\or8\or9\or A\or B\or
C\or D\or E\or F\fi}

\font\twelveeuf=eufm10 scaled 1200

\font\seveneuf=eufm7
\font\fiveeuf=eufm5
\newfam\euffam
\textfont\euffam=\twelveeuf
\scriptfont\euffam=\seveneuf
\scriptscriptfont\euffam=\fiveeuf
\def\euf{\ifcase\euffam 0\or1\or2\or3\or4\or5\or6\or7\or8\or9\or A\or B\or
C\or D\or E\or F\fi}

\def\frak#1{\fam\euffam#1}

\def\Bbb#1{\fam\msbfam#1}

\mathchardef\gapprox"3\msa26
\mathchardef\lapprox"3\msa2E

\begin{document}

\title{ELECTRIC-MAGNETIC DUALITY ROTATIONS IN NON-LINEAR ELECTRODYNAMICS}

\author{G W GIBBONS\\ \&\\D A RASHEED\\ \\D.A.M.T.P.\\University of
Cambridge\\Silver Street\\Cambridge CB3 9EW\\U.K.}

\maketitle

\begin{abstract}
\noindent
We show that there is a function of one variable's worth of  Lagrangians
for a single Maxwell field coupled to gravity whose equations of motion
admit electric-magnetic duality. Such Lagrangians are given by solutions
of the Hamilton-Jacobi equation for timelike geodesics in Witten's
two-dimensional black hole. Among them are the Born-Infeld Lagrangian
which arises in open string theory. We investigate the effect of the
axion and the dilaton in the open superstring case and we show that this
theory loses its electric-magnetic duality invariance when one considers
the higher order electromagnetic field terms. We discuss some
implications for black holes in string theory and an extension to
$2k$-forms in $4k$ spacetime dimensions.
\end{abstract}

\renewcommand{\thepage}{ }
\pagebreak

\renewcommand{\thepage}{\arabic{page}}
\setcounter{page}{1}

\sect{Introduction}

In ordinary Maxwell theory in Minkowski spacetime a Hodge duality
rotation is an action of $SO(2)$
\be
\left\{
\begin{array}{rcl}
{\bf E} & \rightarrow & \cos\alpha ~{\bf E} - \sin\alpha ~{\bf B} \\
 & & \\
{\bf B} & \rightarrow & \cos\alpha ~{\bf B} + \sin\alpha ~{\bf E}
\end{array}
\right.
\ee
which takes solutions of the sourceless Maxwell's equations into
solutions and which moreover commutes with Lorentz transformations. If
one writes the duality transformation in the form
\be
F_{\mu\nu} \rightarrow \cos\alpha ~F_{\mu\nu} + \sin\alpha ~\star F
_{\mu \nu}
\ee
where $\star$ denotes the Hodge star operation one sees that invariance
under duality rotations continues to be a symmetry of Maxwell's equations
in a  curved spacetime. One can ask whether a generalization of duality
invariance continues to hold if one modifies the Maxwell action. In
particular one may consider Lagrangian densities ${\frak L}= \sqrt {-g}
L$ depending only on a single  Maxwell field $F_{\mu\nu} = \partial_\mu
A_\nu - \partial_\nu A_\mu$ and the spacetime metric $g_{\alpha\beta}$
but which are not quadratic in the two-form $F_{\mu \nu}$ and  whose
equations are therefore non-linear. The purpose of this paper is to
answer this question. We shall show that there are as many such
Lagrangians as there are solutions of the Hamilton-Jacobi equation for
timelike geodesics in a two dimensional Witten black hole spacetime
\cite{Wit}. Thus, roughly speaking, such Lagrangians depend upon an
arbitrary function of a single real variable. A particular example is the
Born-Infeld Lagrangian \cite{BorInf}
\be
{\frak L} = {1\over b^2} \left\{\sqrt{-g} - \sqrt{-{\rm det}\left(g_{\mu
\nu} + bF_{\mu\nu}\right)}\right\},
\label{LBID}
\ee
where the constant $b$ has the dimensions of length squared.\footnote{In
this paper we use units in which $c=\hbar=\varepsilon_0=\mu_0=1$, so that
the parameter $b$, Newton's constant $G$ and the inverse string tension
$\alpha^\prime$ all have the dimensions of length squared. In these
units, electric and magnetic charges are dimensionless.} Of course if
one keeps the Lagrangian quadratic in $F_{\mu \nu}$ but  couples extra
fields, such as an axion and dilaton, it is  known \cite{ZumGai} that one
may extend the $SO(2)$ invariance to an $SL(2, {\Bbb R})$ invariance but
as far as we are aware all existing discussions of duality do not
consider equations of motion which are non-linear in the Maxwell field.
Because this case arises in open string theory \cite{FraTse}, with
\be
b=2\pi\alpha^\prime,
\label{alphaprime}
\ee
we consider it worth a special investigation. In particular the
Born-Infeld Lagrangian arises in open superstring theory, together with
an axion and a dilaton, and we shall show later that, in this case, all
electric-magnetic duality is lost.

In what follows we shall adopt the following conventions, $\eta _{\mu \nu
\lambda \rho}$ will be the covariantly constant volume form and indices
will be freely raised and lowered using the metric $g_{\mu\nu}$ whose
signature is $-+++$. Thus the Hodge star operation  is given by $F_{\mu
\nu} \rightarrow \star F_{\mu \nu }= { 1\over 2} \eta _{\mu \nu } ~ ^
{\alpha \beta } F_{\alpha \beta}$ and we have $\star \star =-1$.
The electric intensity $\bf E$ and magnetic induction $\bf B$ are defined
in a local orthonormal frame by $E_i=F_{i0}$ and $B_i= {1 \over 2}
\epsilon _{ijk} F_{jk}$. One has $\star {\bf E} =-{\bf B}$ and $\star
{\bf B} = {\bf E}$. Therefore $F_{\alpha \beta } F^{\alpha \beta} =
2\bigl ({\bf B} ^2- {\bf E}^2 \bigr )$ and $F_{\alpha \beta } \star
F^{\alpha \beta}= 4 {\bf E} \cdot {\bf B}$.  The Bianchi identities are
\be
\partial_{[\alpha} F_{\beta\gamma]} = 0,
\ee
which are equivalent to
\ba
{\bf\nabla}\cdot{\bf B} & = & 0 \nonumber \\
 & & \\
{\bf\nabla}\times{\bf E} & = & - {\partial{\bf B}\over\partial t}.
\nonumber
\ea
Given a Lagrangian $L$ one may define $G^{\mu\nu}$ by \footnote{There is
some ambiguity in the definition of this partial derivative depending on
whether or not one takes into account the antisymmetry of $F_{\mu\nu}$.
Here we treat $F_{\mu\nu}$ and $F_{\nu\mu}$ as independent variables,
hence the factor of 2.}
\be
G^{\mu\nu} = -2 {\partial L\over\partial F_{\mu\nu}}.
\ee
The field equations are
\be
\partial_{[\alpha}\star G_{\beta\gamma]} = 0.
\ee
If one defines the electric induction $\bf D$ and magnetic intensity $\bf
H$ by $ D_i=G_{i0}$ and $ H_i= {1 \over 2} \epsilon _{ijk} G_{jk}$, then
the field equations are equivalent in Minkowski spacetime to
\ba
{\bf\nabla}\cdot{\bf D} & = & 0 \nonumber \\
 & & \\
{\bf\nabla}\times{\bf H} & = & + {\partial{\bf D}\over\partial t}.
\nonumber
\ea
The field equations and  Bianchi identities may be combined in the form
\ba
{\bf\nabla} \cdot ({\bf D} + i {\bf B}) & = & 0 \nonumber \\
 & & \\
{\bf\nabla} \times ({\bf E} + i {\bf H}) & = & i{\partial\over\partial t}
({\bf D} + i {\bf B}). \nonumber
\ea
Thus the generally covariant generalization of Hodge duality rotations to
a general electric-magnetic duality rotation through an angle $\alpha$ is
\be
\left\{
\begin{array}{rcl}
F_{\mu\nu} & \rightarrow & \cos\alpha ~F_{\mu\nu} + \sin\alpha ~\star
G_{\mu\nu} \\
 & & \\
G_{\mu\nu} & \rightarrow & \cos\alpha ~G_{\mu\nu} + \sin\alpha ~\star
F_{\mu \nu}.
\end{array}
\right.
\label{Rot}
\ee
In Minkowski spacetime this is equivalent to
\be
\left\{
\begin{array}{rcl}
{\bf E} + i {\bf H} & \rightarrow & e^{i\alpha} ({\bf E} + i{\bf H}) \\
 & & \\
{\bf D} + i {\bf B} & \rightarrow & e^{i\alpha} ({\bf D} + i {\bf B}).
\end{array}
\right.
\ee
It is clear that if $F_{\mu \nu}$ and $G_ {\mu \nu}$ were independent
variables then  electric-magnetic duality rotations would take the linear
field equations into the linear Bianchi identities and conversely
regardless of the specific form of the Lagrangian $L$. However they are
not independent of one another; they are linked by the constitutive
relation $G^{\mu \nu} = -2 {\partial L \over \partial F_{\mu \nu} }$
which in general is a non-linear relation between $G_{\mu \nu}$ and
$F_{\mu \nu}$. Only for a restricted class of Lagrangians will this
relation be invariant under electric-magnetic duality rotations. Note
that one should distinguish electric-magnetic duality rotations from
Hodge rotations. Hodge rotations  transform  $\bf E$ into $\bf B$ and
$\bf D$ into $\bf H$. Except for the case of Maxwell theory, for
which  $F_{\mu \nu}= G_{\mu \nu}$ and hence ${\bf E}={\bf D}$ and $ {\bf
B}= {\bf H}$, Hodge rotations do not take the Bianchi identities into the
equations of motion.

\sect{Duality Invariance}

In order to find which Lagrangians admit electric-magnetic duality
rotations it  suffices to consider infinitesimal transformations which
take the form
\be
\left\{
\begin{array}{rcl}
\delta F_{\mu\nu} & = & \star G_{\mu\nu} \\
 & & \\
\delta G_{\mu\nu} & = & \star F_{\mu\nu}.
\end{array}
\right.
\ee
The invariance of the constitutive relation under an infinitesimal
duality rotation requires that
\be
{1\over 2} \eta^{\mu\nu\lambda\rho} F_{\lambda\rho} =
{1\over 2} \eta_{\sigma\tau\alpha\beta} G^{\alpha\beta}
{\partial\over\partial F_{\sigma\tau}}
\left(-2{\partial L\over\partial F_{\mu\nu}}\right).
\ee
Substituting the definition of $G_{\alpha \beta}$  and using the
commutation of partial derivatives gives
\be
{1\over 2} \eta^{\mu\nu\lambda\rho} F_{\lambda\rho} =
2 \eta_{\sigma\tau\alpha\beta} {\partial L\over\partial F_{\alpha\beta}}
{\partial\over\partial F_{\mu\nu}}
\left({\partial L\over\partial F_{\sigma\tau}}\right).
\ee
This second order  partial differential equation in the six variables
$F_{\mu \nu}$ is the necessary and sufficient condition on the Lagrangian
$L$ that its Euler-Lagrange equations admit duality invariance. Because
$\eta _{\alpha \beta \sigma \tau}=\eta _{ \sigma \tau \alpha \beta}$ we
have
\be
{1\over 2} \eta^{\mu\nu\lambda\rho} F_{\lambda\rho} =
{\partial\over\partial F_{\mu\nu}} \left(\eta_{\sigma\tau\alpha\beta}
{\partial L\over\partial F_{\sigma\tau}} {\partial L\over\partial
F_{\alpha\beta}}\right).
\ee
A first integral of our second  order partial differential equation is
obtained by integrating with respect to $F_{\mu \nu}$, remembering again
that $\eta _{\alpha \beta \sigma \tau}=\eta _{ \sigma \tau \alpha
\beta}$. This gives the first order partial differential equation~:
\be
{1\over 4} \eta^{\mu\nu\lambda\rho} F_{\mu\nu} F_{\lambda\rho} =
\eta_{\sigma\tau\alpha\beta} {\partial L\over\partial F_{\sigma\tau}}
{\partial L\over\partial F_{\alpha\beta}} + 2C,
\label{FirstO}
\ee
where $C$ is an arbitrary constant of integration. In fact, if $L$ is to
agree with the usual Maxwell Lagrangian at weak fields, the constant must
vanish but we shall retain it for the time being. Note that the condition
we have obtained is manifestly Lorentz-invariant. By differentiating the
first integral one sees that every such first integral satisfies the
second order equation provided the constitutive relation holds.

Using the definition of $G_{\mu \nu}$ we may write the first integral as
\be
F_{\mu\nu} \star F^{\mu\nu} = G_{\mu\nu} \star G^{\mu\nu} + 4C.
\label{FirstI}
\ee
In terms of the electric and magnetic intensities $\bf E$ and $\bf H$ and
the electric and magnetic inductions $\bf D$ and $\bf B$ our condition
reads
\be
{\bf E} \cdot {\bf B}= {\bf D} \cdot {\bf H} + C.
\ee

Equation (\ref{FirstI}) has an important consequence for the duality
transformation properties of the energy-momentum tensor, which is defined
by
\be
\sqrt{-g}~T_{\mu\nu} = -2 {\partial\over\partial g^{\mu\nu}}
\left(\sqrt{-g}~L\right),
\ee
where the partial derivative with respect to $g_{\mu\nu}$ is taken with
$F_{\mu\nu}$ held fixed. Under an infinitesimal duality rotation
$T_{\mu\nu}$ will therefore transform according to
\ba
\sqrt{-g}~\delta T_{\mu\nu} & = & -2 {\partial\over\partial g^{\mu\nu}}
\left(\sqrt{-g}~\delta L\right) \nonumber \\
\nonumber \\
 & = & -2 {\partial\over\partial g^{\mu\nu}} \left(\sqrt{-g}~
{\partial L\over\partial F_{\lambda\rho}} \star
G_{\lambda\rho}\right) \\
\nonumber \\
 & = &  {\partial\over\partial g^{\mu\nu}} \left(\sqrt{-g}~
G^{\lambda\rho}\star G_{\lambda\rho}\right). \nonumber
\ea
If equation (\ref{FirstI}) holds then
\ba
\sqrt{-g}~\delta T_{\mu\nu} & = & {\partial\over\partial g^{\mu\nu}}
\left(\sqrt{-g}~F^{\lambda\rho}\star F_{\lambda\rho} - 4C\sqrt{-g}\right)
\nonumber \\
\nonumber \\
 & = & {1\over 2}F_{\lambda\rho}F_{\sigma\kappa} {\partial\over\partial
g^{\mu\nu}} \left(\sqrt{-g}~\eta^{\lambda\rho\sigma\kappa}\right) +
2Cg_{\mu\nu}\sqrt{-g} \\
\nonumber \\
 & = & 2Cg_{\mu\nu}\sqrt{-g}. \nonumber
\ea
Thus the energy-momentum tensor of a duality invariant theory with $C=0$
is itself invariant under duality rotations.

\sect{Hamiltonian Viewpoint}

The condition that we have obtained has an interesting
Hamiltonian geometric interpretation which is  useful in discussing its
solution and which also applies to more complicated models. The
six-dimensional space $V=\Lambda ^2 ({\Bbb R}^4)$ of two-forms in ${\Bbb
R}^4$,  has coordinates $F_{\mu \nu}$ and carries a Lorentz-invariant
metric $k$ with signature $+++---$ defined by
\be
k(F,F) = {1\over 2} \eta^{\alpha\beta\mu\nu} F_{\alpha\beta} F_{\mu\nu} =
4 {\bf E} \cdot {\bf B}.
\ee
The dual space $V^*$ of $V$ consists of skew-symmetric second rank
contravariant tensors $G^{\mu \nu}$. The phase space ${\cal P}=V\oplus
V^*$ carries a natural symplectic structure~:
\be
{1\over 2} dG^{\mu\nu} \wedge dF_{\mu\nu} = d{\bf B} \wedge\cdot d{\bf H}
- d{\bf D} \wedge\cdot d{\bf E}.
\ee
The constitutive relation connecting $G^{\mu \nu}$ and $F_{\mu \nu}$ is
a Legendre transformation with generating function $L$ and defines
a Lagrangian submanifold $\cal L$ of the phase space ${\cal P}=V\oplus
V^*$.

Electric-magnetic duality rotations act symplectically on ${\cal P}$
with generating function or moment map $K: {\cal P}\rightarrow{\Bbb R}$
given by
\be
K = {1\over 2} \left(G_{\mu\nu} \star G^{\mu\nu} - F_{\mu\nu} \star
F^{\mu\nu}\right).
\ee
The Legendre transformation will therefore commute with electric-magnetic
duality rotations if the Lagrangian submanifold $\cal L$ is invariant
under the action of electric-magnetic duality rotations.

Now it is true quite generally that a Lagrangian submanifold
$p_i={\partial S \over \partial q^i}$, with generating function
$S(q^i)$, of a symplectic manifold ${\cal P}$ with canonical coordinates
$p_i,q^i$, is invariant under the Hamiltonian flow generated by a moment
map $K(p_i,q^i)$ if and only if $S$ satisfies the Hamilton-Jacobi
equation associated to the Hamiltonian function $K(p_i,q^i)$, that is
\be
K\left({\partial S\over\partial q^i},q^i\right) = \kappa,
\ee
where $\kappa$ is a constant labelling the level sets of the Hamiltonian
$K$ on which the flow lies. To obtain the present case we set
$S(q^i)=L( F_{\mu \nu})$ and $\kappa=-2C$.

One may, if one wishes, regard $L$ as satisfying the Hamilton-Jacobi
equation associated to  geodesic flow with respect to the  Jacobi-metric
$h$ on $V=\Lambda ^2({\Bbb R}^4)$ given by
\be
h(F,F) = {1\over 2} {dF_{\mu\nu} \otimes \star dF^{\mu\nu} \over
F_{\alpha\beta} \star F^{\alpha\beta} - 4C}.
\ee

For a Lorentz-invariant theory the Lagrangian function $L=L(F_{\mu \nu})$
can only depend on the two independent Lorentz scalars $F_{\alpha \beta }
F^{\alpha \beta} = 2\bigl ({\bf B} ^2- {\bf E}^2 \bigr )$  and $F_{\alpha
\beta } \star F^{\alpha \beta}= 4 {\bf E} \cdot {\bf B}$.
Thus our condition effectively reduces to a partial differential equation
in two variables. It turns out to be most convenient to
make use of the freedom to make Lorentz-transformations to pass to a
frame in which $\bf E$ and $\bf B$ , are parallel. It then follows that
both $\bf D$ and $\bf H$ will also be parallel to both $\bf E$ and $\bf
B$. This is possible everywhere in $V$ except on  the ``lightcone''  of
the metric $k$ where the invariant $F_{\alpha\beta} \star F^{\alpha
\beta}=4{\bf E}\cdot{\bf B}$ vanishes. We therefore introduce the two
variables given by
\ba
E^2 & = & {1\over 2} \left\{{\bf E}^2 - {\bf B}^2 + \sqrt{\left({\bf B}^2
- {\bf E}^2\right)^2 + 4\left({\bf E} \cdot {\bf B}\right)^2}\right\}
\nonumber \\
\\
B^2 & = & {1\over 2} \left\{{\bf B}^2 - {\bf E}^2 + \sqrt{\left({\bf B}^2
- {\bf E}^2\right)^2 + 4\left({\bf E} \cdot {\bf B}\right)^2}\right\}
\nonumber
\ea
so that $E$ and $B$ are the magnitudes of  $\bf E$ and $\bf B$ in this
special frame. In terms of $E$ and $B$ the Hamilton-Jacobi equation
becomes
\be
{\partial L\over\partial E} {\partial L\over\partial B} = C - EB.
\label{DEEB}
\ee
This is the Hamilton-Jacobi equation for geodesics in the two-dimensional
metric
\be
ds^2 = {dE dB \over C-EB}.
\label{Witmet}
\ee
Interestingly $E$ and $B$ are null coordinates for Witten's two
dimensional black hole spacetime $W$ \cite{Wit}. The general solution of
this equation may be constructed by picking an initial spacelike curve
$\Sigma$ in $W$ on which $L$ takes some constant  value. The value of $L$
at a point in $p= (E,B)~ \in W$ not on   $\Sigma$ is then the action or
proper time along a timelike geodesic joining $p$ to $\Sigma$. Since
there may be more than one such geodesic the Lagrangian $L$ will in
general have branch points and be multi-valued. The initial curve
$\Sigma$ may be specified by giving an arbitrary function of a single
variable. Alternatively one may fix a Cauchy  curve $\Sigma$ on which $L$
is allowed to take arbitrary values. $L$ is then obtained at a point $p
\in W$ not on  $\Sigma$ by joining $p$ to $\Sigma$ by a timelike geodesic
of proper time $\tau(p)$ meeting $\Sigma$ in the point $q(p)$. One then
has
\be
L(p)=L(q(p))+\tau(p).
\ee
Convenient initial Cauchy surfaces are the null surfaces $B=0$ or $E=0$.
Thus for example one may specify arbitrarily the constitutive relation
$D={\partial L \over \partial E}$ at zero magnetic field or $H=-{\partial
L \over \partial B}$ at zero electric field and extend it to
non-vanishing magnetic or electric fields respectively in such a way that
it is duality invariant. Thus  there are, roughly speaking, as many
electric-magnetic duality invariant generalizations of Maxwell's
equations as there are functions of a single variable. Physically one
presumably  wants them to coincide with Maxwell's theory at small values
of $E$ and $B$ in which case a necessary condition is that $C=0$. If
$C=0$ and
\be
u = {1\over 2} E^2 \quad ,\quad v = {1\over 2} B^2,
\ee
then the  Hamilton-Jacobi equation becomes even simpler; it reduces to
that for timelike geodesics in two-dimensional Minkowski spacetime
\be
{\partial L\over\partial u} {\partial L\over\partial v} = -1.
\label{DEuv}
\ee

\sect{Duality Invariant Lagrangians}

To obtain explicit solutions of (\ref{DEuv}) one must resort to
techniques such as separation of variables in particular coordinate
systems. For example if one supposes that the solution separates
multiplicatively in $(u,v)$ coordinates one obtains
\be
L = \pm \sqrt{\alpha-\beta E^2} \sqrt{\gamma-\delta B^2} \quad ,\quad
\beta\delta = 1.
\label{SepL}
\ee
For general $\alpha,\beta,\gamma,\delta$ this does not coincide with
Maxwell theory for small values of $E$ and $B$, the terms quadratic in $E$
and $B$ being different from those in Maxwell theory, so the
condition $C=0$ is not sufficient to ensure that the theory has the
correct weak field limit. It can be shown that the only Lagrangian of the
form (\ref{SepL}) which does have the correct weak field limit is given by
$\delta=1/\beta$, $\gamma=\alpha/\beta^2$, $\beta=\alpha b^2$ which is
equivalent to the Born-Infeld Lagrangian
\ba
L & = & {1\over b^2} \left\{1 - \sqrt{1 + b^2 \left({\bf B}^2 - {\bf
E}^2\right) - b^4 \left({\bf E} \cdot {\bf B}\right)^2}\right\}
\nonumber \\
 & & \label{LBI} \\
 & = & {1\over 2}\left({\bf E}^2-{\bf B}^2\right) + {1\over
8}b^2\left({\bf E}^2-{\bf B}^2\right)^2 + {1\over 2}b^2\left({\bf
E}\cdot{\bf B}\right)^2 + {\cal O}(6), \nonumber
\ea
where ${\cal O}(6)$ denotes terms of order 6 in ${\bf E}$ and ${\bf B}$.
The fact that Born-Infeld theory is electric-magnetic duality invariant
seems to have been noticed first by Schr\"odinger \cite{Schro}. It is
interesting to note that the Euler-Heisenberg Lagrangian \cite{EulHei}
for a supersymmetric system of minimally coupled spin-${1\over 2}$ and
spin-$0$ particles agrees with the Born-Infeld Lagrangian, with
\be
b^2={e^4\over 24\pi^2m^4},
\label{Euler}
\ee
up to and including terms ${\cal O}(4)$, and
hence it is duality invariant to that order. This may indicate that,
under some circumstances, classical duality may persist at the quantum
level.

Approximate solutions to equation (\ref{DEuv}) can also be obtained by a
power series expansion. It is convenient to define new
variables $x$ and $y$ by
\ba
x = & u+v & = {1\over 2}\left(E^2+B^2\right) \nonumber \\
 & & \\
y = & u-v & = {1\over 2}\left(E^2-B^2\right). \nonumber
\ea
Then (\ref{DEuv}) becomes
\be
\left({\partial L\over\partial x}\right)^2 - \left({\partial
L\over\partial y}\right)^2 = -1.
\label{DExy}
\ee
Assuming that $L$ coincides with the Maxwell Lagrangian for small values
of $E$ and $B$, i.e.\ $L=y+{\cal O}(x^2,y^2)$, then it is found that the
quadratic terms in $x$ and $y$ must be proportional to $x^2$, i.e.
\ba
L & = & y + ax^2 + {\cal O}(x^3,y^3) \nonumber \\
 & & \\
 & = & {1\over 2}\left({\bf E}^2-{\bf B}^2\right) + {1\over
4}a\left({\bf E}^2-{\bf B}^2\right)^2 + a\left({\bf
E}\cdot{\bf B}\right)^2 + {\cal O}(6). \nonumber
\ea
This coincides with the Born-Infeld Lagrangian up to and including terms
quartic in ${\bf E}$ and ${\bf B}$, on setting $a={1\over 2}b^2$.

Equation (\ref{DExy}) is insufficient to determine uniquely the higher
order terms in the expansion of $L$. Writing $L$ in the form
\be
L = y + f(x) + \sum_{n=3}^\infty L^{(n)}(x,y),
\ee
where $f$ is an arbitrary function of $x$ satisfying $f(0)=f^\prime(0)=0$
and each $L^{(n)}(x,y)$ contains only terms of order $n$ in $x$ and $y$,
equation (\ref{DExy}) determines each function $L^{(n)}(x,y)$ uniquely in
terms of $f(x)$~:
\ba
L^{(3)} & = & {1\over 2}y\left(f^\prime(x)\right)^2 \nonumber \\
 & & \\
L^{(n+1)} & = & \int dy\left\{ 1 - \left(1+{\partial\over\partial
y}\sum_{m=3}^n L^{(m)}\right)^2 + \left(f^\prime(x)+{\partial\over\partial
x}\sum_{m=3}^n L^{(m)}\right)^2 \right\}. \nonumber \\
 & & {\rm (Keeping\ only\ terms\ of\ order\ }n+1{\rm \ in\ }x,y).
\nonumber
\ea
This confirms the fact that there are a function of one variable's worth
of Lagrangians admitting duality rotations and gives an explicit
algorithm for their construction. This method of construction guarantees
that
\be
L = y + f(x) + \sum_{m=3}^n L^{(m)}(x,y)
\ee
satisfies (\ref{DExy}) up to and including terms of order $n-1$ in $x$
and $y$ and hence leads to dual invariant equations of motion to that
order.

Exact solutions of the differential equation (\ref{DEEB}) are, in
practice, hard to find. One such solution, with $C\ne 0$, may be obtained
by separation of variables, using the variables $E+B$ and $E-B$. This
procedure gives
\ba
L & = & {1\over 2}K\left(B+E\right)\sqrt{1-{\left(B+E\right)^2\over 4K^2}}
 + K^2\sin^{-1}\left({B+E\over 2K}\right) \nonumber \\
 & + & {1\over
2}\sqrt{K^2-C}\left(B-E\right)\sqrt{1-{\left(B-E\right)^2\over
4\left(K^2-C\right)}} \\
 & + & \left(K^2-C\right)\sin^{-1}\left({B-E\over 2\sqrt{K^2-C}}\right),
\nonumber
\ea
where $K$ is an arbitrary constant. However, this solution is
pathological since the ${\bf D}$ and ${\bf H}$ fields defined by
\be
{\bf D} = {\partial L\over\partial{\bf E}}
\quad,\quad
{\bf H} = - {\partial L\over\partial{\bf B}}
\ee
are ill-defined at ${\bf E}={\bf B}=0$, since the limit of ${\bf D}$ and
${\bf H}$ as ${\bf E}$ and ${\bf B}$ are taken to zero is dependant on
the way in which the limit is taken.

\sect{Other Dualities}

One may ask if there are Lagrangians leading to other more general
electric-magnetic dualities. The $SO(2)$ rotation (\ref{Rot}) may be
generalized to an $SL(2,{\Bbb R})$ duality transformation.
Infinitesimally this takes the form
\be
\left\{
\begin{array}{rcl}
\delta F_{\mu\nu} & = & \alpha F_{\mu\nu} + \beta\star G_{\mu\nu} \\
 & & \\
\delta G_{\mu\nu} & = & -\alpha G_{\mu\nu} - \gamma\star F_{\mu\nu}.
\end{array}
\right.
\label{SL2R1}
\ee
In terms of ${\bf E}$, ${\bf B}$, ${\bf D}$ and ${\bf H}$ this gives
\be
\left\{
\begin{array}{rcl}
\delta{\bf E} & = & \alpha{\bf E} - \beta{\bf H} \\
 & & \\
\delta{\bf B} & = & \alpha{\bf B} + \beta{\bf D} \\
 & & \\
\delta{\bf D} & = & -\alpha{\bf D} - \gamma{\bf B} \\
 & & \\
\delta{\bf H} & = & -\alpha{\bf H} + \gamma{\bf E}.
\end{array}
\right.
\label{SL2R2}
\ee
As before, one can work in a Lorentz frame in which ${\bf E}$, ${\bf B}$,
${\bf D}$ and ${\bf H}$ are all parallel. The invariance of the
constitutive relation then gives a pair of second order partial
differential constraints on $L$~:
\ba
{\partial\over\partial E} \left(\alpha E{\partial L\over\partial E} +
\beta{\partial L\over\partial E}{\partial L\over\partial B}\right) +
\alpha B{\partial^2 L\over\partial E\partial B} + \gamma B & = & 0
\nonumber \\
 & & \label{DiffCon} \\
{\partial\over\partial B} \left(\alpha B{\partial L\over\partial B} +
\beta{\partial L\over\partial B}{\partial L\over\partial E}\right) +
\alpha E{\partial^2 L\over\partial E\partial B} + \gamma E & = & 0.
\nonumber
\ea
However, if one requires that $L$ coincides with Maxwell theory up to
and including terms quadratic in $E$ and $B$, then by expanding $L$ as a
power series in $E$ and $B$ and considering terms linear in $E$ and $B$
in equation (\ref{DiffCon}), it can be shown that no solutions exist
unless $\alpha=0$ and $\beta=\gamma$. The duality transformation then
reduces to the $SO(2)$ rotations considered above.

Therefore, restricted to Lagrangians which coincide with the Maxwell
Lagrangian for small values of ${\bf E}$ and ${\bf B}$, the only
continuous electric-magnetic duality transformations which leave the
constitutive relation invariant are $SO(2)$ rotations of the form
(\ref{Rot}). Lagrangians which give rise to more general $SL(2,{\Bbb R})$
dualities cannot coincide with the Maxwell Lagrangian for small ${\bf E}$
and ${\bf B}$ and must be of the form
\be
L = \lambda \left({\bf E}^2-{\bf B}^2\right) + \theta{\bf E}\cdot{\bf B}
+ {\cal O}(4),
\ee
since there is no invariant of $F_{\mu\nu}$ of odd order in the fields.
However $\lambda$ can be made equal to ${1\over 2}$ and the $\theta$-term
made to vanish by a canonical coordinate transformation (rescaling ${\bf
E}$ and ${\bf B}$ and rotating ${\bf E}$ into ${\bf B}$). If the
Maxwell field is the only field present, it therefore suffices to
consider Lagrangians of the form
\be
L = {1\over 2} \left({\bf E}^2-{\bf B}^2\right) + {\cal O}(4)
\ee
and duality transformations of the form (\ref{Rot}).

Alternatively, the duality transformation properties of Lagrangians with
$\theta$-terms may be described by a conjugated action of an $SO(2)$
rotation. Consider a Lagrangian ${\widetilde L}$ of the form
\be
{\widetilde L} = L + \theta{\bf E}\cdot{\bf B}
\ee
where the Lagrangian $L$ admits an ordinary electric-magnetic duality
rotation, i.e.\ ${\partial L\over\partial E}{\partial L\over\partial
B}=-BE$. For which values of $\alpha,\beta,\gamma$ (if any) does
${\widetilde L}$ admit an $SL(2,{\Bbb R})$ duality transformation of the
form (\ref{SL2R1}), (\ref{SL2R2})? This is equivalent to asking, for
which values of $\alpha,\beta,\gamma$ does ${\widetilde L}$ satisfy
equations (\ref{DiffCon})? It is easy to check that, in general,
${\widetilde L}$ will only satisfy (\ref{DiffCon}) if
$\alpha=-\theta\beta$ and $\gamma=\beta\left(1+\theta^2\right)$. The
resulting duality may be rewritten as~:
\be
\left({{\bf D}\atop{\bf B}}\right) \rightarrow S \left({{\bf D}\atop{\bf
B}}\right) \quad,\quad \left({{\bf E}\atop{\bf H}}\right) \rightarrow
\left(S^T\right)^{-1} \left({{\bf E}\atop{\bf H}}\right) \quad,\quad
{\cal M} \rightarrow \left(S^T\right)^{-1}{\cal M}S^{-1}
\ee
where ${\cal M({\bf B},{\bf D})}$ is the matrix which encodes the
constitutive relation~:
\be
\left({{\bf E}\atop{\bf H}}\right) = {\cal M} \left({{\bf D}\atop{\bf
B}}\right)
\ee
and
\be
S = \left(
\begin{array}{cc}
1+\theta\alpha & -\alpha\left(1+\theta^2\right) \\
\alpha & 1-\theta\alpha
\end{array}
\right).
\ee
This duality is a conjugate action of an $SO(2)$ rotation. This may be
seen by writing $S$ in the form
\be
S = ARA^{-1}
\ee
where
\be
A = \left(
\begin{array}{cc}
1 & \theta \\
0 & 1
\end{array}
\right)
\label{A}
\ee
and $R$ is an infinitesimal rotation~:
\be
R = \left(
\begin{array}{cc}
1 & -\alpha \\
\alpha & 1
\end{array}
\right).
\ee
The full group action may then be reconstructed simply by taking $R$ to
be a finite rotation~:
\be
R = \left(
\begin{array}{cc}
\cos\alpha & -\sin\alpha \\
\sin\alpha & \cos\alpha
\end{array}
\right) \quad,\quad
S = \left(
\begin{array}{cc}
\cos\alpha+\theta\sin\alpha & -\left(1+\theta^2\right)\sin\alpha \\
\sin\alpha & \cos\alpha-\theta\sin\alpha
\end{array}
\right).
\ee
Thus adding a $\theta$-term to a Lagrangian which admits an $SO(2)$
duality rotation merely conjugates the rotation by a matrix $A$ of the
form (\ref{A}).

It is also interesting to ask whether a particular Lagrangian leads to a
theory with a {\em discrete} electric-magnetic duality invariance. In the
case of Born-Infeld theory, given by the Lagrangian (\ref{LBI}), the
constitutive relation gives ${\bf D}$ and ${\bf H}$ in terms of ${\bf E}$
and ${\bf B}$~:
\ba
{\bf D} = + {\partial L\over\partial{\bf E}} & = & {{\bf E} +
b^2\left({\bf E}\cdot{\bf B}\right){\bf B}\over\sqrt{1 + b^2\left({\bf
B}^2-{\bf E}^2\right) - b^4\left({\bf E}\cdot{\bf B}\right)^2}} \nonumber
\\
 & & \\
{\bf H} = - {\partial L\over\partial{\bf B}} & = & {{\bf B} -
b^2\left({\bf E}\cdot{\bf B}\right){\bf E}\over\sqrt{1 + b^2\left({\bf
B}^2-{\bf E}^2\right) - b^4\left({\bf E}\cdot{\bf B}\right)^2}}. \nonumber
\ea
These equations can be solved to give ${\bf E}$ and ${\bf H}$ in terms of
${\bf D}$ and ${\bf B}$~:
\ba
{\bf E} & = & {\left(1+b^2{\bf B}^2\right){\bf D} - b^2\left({\bf
B}\cdot{\bf D}\right){\bf B}\over\sqrt{\left(1+b^2{\bf
B}^2\right)\left(1+b^2{\bf D}^2\right) - b^4\left({\bf B}\cdot{\bf
D}\right)^2}} \nonumber \\
 & & \label{EHDB} \\
{\bf H} & = & {\left(1+b^2{\bf D}^2\right){\bf B} - b^2\left({\bf
B}\cdot{\bf D}\right){\bf D}\over\sqrt{\left(1+b^2{\bf
B}^2\right)\left(1+b^2{\bf D}^2\right) - b^4\left({\bf B}\cdot{\bf
D}\right)^2}}. \nonumber
\ea
When expressed in this form, it is clear that there is a discrete
electric-magnetic duality corresponding to interchanging ${\bf B}$ and
${\bf D}$ and also interchanging ${\bf E}$ and ${\bf H}$. In the
Hamiltonian picture, this discrete duality invariance of the constitutive
relation may be seen as a mirror symmetry of the Lagrangian submanifold
${\cal L}$ under reflections in the 6-dimensional hyper-plane ${\bf
B}={\bf D}$, ${\bf E}={\bf H}$ of the phase space ${\cal P}$.

To determine whether other Lagrangians also admit this discrete duality
it is sufficient to consider a power series expansion of the Lagrangian.
Assuming that the Lagrangian coincides with the Maxwell Lagrangian for
weak fields, the power series will take the form
\ba
L & = & {1\over 2}\left({\bf E}^2-{\bf B}^2\right) + \alpha\left({\bf
E}^2-{\bf B}^2\right)^2 + \beta\left({\bf E}\cdot{\bf B}\right)^2
\nonumber \\
 & & \\
& + & \gamma\left({\bf E}^2-{\bf B}^2\right)^3 + \delta\left({\bf
E}^2-{\bf B}^2\right)\left({\bf E}\cdot{\bf B}\right)^2 + {\cal O}(8).
\nonumber
\ea
Differentiating with respect to ${\bf E}$ and ${\bf B}$ to obtain ${\bf
D}$ and ${\bf H}$ one can then solve for ${\bf E}$ and ${\bf H}$ in terms
of ${\bf B}$ and ${\bf D}$ as above. Then the condition that the
constitutive relation be invariant under discrete duality transformations
gives the following constraints on the coefficients in the power series
expansion of $L$~:
\be
\gamma = 4\alpha^2 \quad ,\quad \delta = 8\alpha\beta - \beta^2.
\label{DisCon}
\ee
It is easy to check that the power series expansion of the Born-Infeld
Lagrangian does indeed satisfy these constraints. On the other hand, the
logarithmic Lagrangian which has recently aroused interest in the theory
of {\sl black points} \cite{Sol},
\be
L = - {1\over 4b^2}\log\left(1+b^2F_{\mu\nu}F^{\mu\nu}\right),
\ee
does not satisfy the constraints (\ref{DisCon}) and so does not have this
discrete duality invariance.

\sect{Extension to $2k$-Forms in $4k$ Spacetime Dimensions}

The analysis above may be repeated in an almost identical fashion
for fields given by $2k$-forms in $4k$ spacetime dimensions where the
operator identity $\star\star=-1$ continues to hold. The Lagrangian will
be a function of the completely antisymmetric tensor field
$F_{\mu_1\dots\mu_{2k}}$ where the $\mu_i$ range from $0$ up to $4k-1$.
$G^{\mu_1\dots\mu_{2k}}$ may be defined by
\be
G^{\mu_1\dots\mu_{2k}} = -(2k)!{\partial L\over\partial
F}_{\mu_1\dots\mu_{2k}}.
\label{Constit}
\ee
In a local orthonormal frame, one may define the electric and magnetic
intensities and inductions in the obvious way~:
\ba
E_{i_1\dots i_{2k-1}} & = & F_{i_1\dots i_{2k-1}0} \nonumber \\
\nonumber \\
B_{i_1\dots i_{2k-1}} & = & {1\over (2k)!}\epsilon_{i_1\dots
i_{2k-1}j_1\dots j_{2k}}F_{j_1\dots j_{2k}} \nonumber\\
\\
D_{i_1\dots i_{2k-1}} & = & G_{i_1\dots i_{2k-1}0} \nonumber \\
\nonumber \\
H_{i_1\dots i_{2k-1}} & = & {1\over (2k)!}\epsilon_{i_1\dots
i_{2k-1}j_1\dots j_{2k}}G_{j_1\dots j_{2k}} \nonumber
\ea
where $i_1,i_2,\dots$ run from 1 up to $4k-1$ and $\epsilon_{i_1\dots
i_{4k-1}}$ is the L\' evi-Civita symbol in $4k-1$ space. The Bianchi
identities are
\be
\partial_{[\lambda}F_{\mu_1\dots\mu_{2k}]} = 0
\ee
which are equivalent to
\ba
{\partial\over\partial x_{i_1}}B_{i_1\dots i_{2k-1}} & = & 0 \nonumber \\
\\
\epsilon_{i_1\dots i_{2k-1}lm_1\dots m_{2k-1}}{\partial\over\partial x_l}
E_{m_1\dots m_{2k-1}} & = & - {\partial\over\partial t}B_{i_1\dots
i_{2k-1}}. \nonumber
\ea
The field equations are
\be
\partial_{[\lambda}\star G_{\mu_1\dots\mu_{2k}]} = 0
\ee
where now the Hodge star operation is defined by
\be
\star G_{\mu_1\dots\mu_{2k}} = {1\over (2k)!}
{\eta_{\mu_1\dots\mu_{2k}}}^{\nu_1\dots\nu_{2k}}G_{\nu_1\dots\nu_{2k}}
\ee
and $\eta_{\mu_1\dots\mu_{4k}}$ is the covariantly constant volume form
in $4k$ dimensional spacetime so that $\star\star=-1$. The field
equations are equivalent to
\ba
{\partial\over\partial x_{i_1}}D_{i_1\dots i_{2k-1}} & = & 0 \nonumber \\
\\
\epsilon_{i_1\dots i_{2k-1}lm_1\dots m_{2k-1}}{\partial\over\partial x_l}
H_{m_1\dots m_{2k-1}} & = & + {\partial\over\partial t}D_{i_1\dots
i_{2k-1}}. \nonumber
\ea
Defining $SO(2)$ electric-magnetic duality rotations as before, invariance
of the constitutive relation (\ref{Constit}) under infinitesimal
transformations~:
\be
\left\{
\begin{array}{rcl}
\delta F & = & \star G \\
 & & \\
\delta G & = & \star F
\end{array}
\right.
\ee
gives the second order differential equation
\be
\eta^{\mu_1\dots\mu_{2k}\nu_1\dots\nu_{2k}} F_{\nu_1\dots\nu_{2k}} =
\eta_{\lambda_1\dots\lambda_{2k}\sigma_1\dots\sigma_{2k}}
G^{\sigma_1\dots\sigma_{2k}} {\partial\over\partial
F}_{\lambda_1\dots\lambda_{2k}} \left( -(2k)! {\partial L\over\partial
F}_{\mu_1\dots\mu_{2k}} \right).
\ee
Rearranging this as before gives \footnote{Note that in $2k$ spacetime
dimensions, for $k$ odd, the right hand side of equation
(\ref{SecondO2k}) vanishes identically, since
$\eta_{\mu_1\dots\mu_k\nu_1\dots\nu_k}=
-\eta_{\nu_1\dots\nu_k\mu_1\dots\mu_k}$ for $k$ odd. Thus duality
rotations of the form (\ref{Rot}) are not possible in 2, 6, 10,
\dots~spacetime
dimensions.}
\ba
{1\over(2k)!} \eta^{\mu_1\dots\mu_{2k}\nu_1\dots\nu_{2k}}
F_{\nu_1\dots\nu_{2k}} \hspace{7cm} \nonumber \\
\hspace{2cm} = {(2k)!\over 2} {\partial\over\partial
F}_{\mu_1\dots\mu_{2k}} \left(
\eta_{\lambda_1\dots\lambda_{2k}\sigma_1\dots\sigma_{2k}} {\partial
L\over\partial F}_{\sigma_1\dots\sigma_{2k}} {\partial L\over\partial
F}_{\lambda_1\dots\lambda_{2k}} \right),
\label{SecondO2k}
\ea
using the fact that
$\eta_{\lambda_1\dots\lambda_{2k}\sigma_1\dots\sigma_{2k}} =
\eta_{\sigma_1\dots\sigma_{2k}\lambda_1\dots\lambda_{2k}}$. This can now
be integrated to give the first order differential equation which
generalizes (\ref{FirstO})~:
\ba
{1\over 2(2k)!} \eta^{\mu_1\dots\mu_{2k}\nu_1\dots\nu_{2k}}
F_{\mu_1\dots\mu_{2k}} F_{\nu_1\dots\nu_{2k}} \hspace{6cm} \nonumber \\
\hspace{3cm} = {(2k)!\over 2}
\eta_{\mu_1\dots\mu_{2k}\nu_1\dots\nu_{2k}} {\partial L\over\partial
F}_{\mu_1\dots\mu_{2k}} {\partial L\over\partial F}_{\nu_1\dots\nu_{2k}}
+ 2C.
\ea
Using the constitutive relation, this is equivalent to
\be
F_{\mu_1\dots\mu_{2k}} \star F^{\mu_1\dots\mu_{2k}} =
G_{\mu_1\dots\mu_{2k}} \star G^{\mu_1\dots\mu_{2k}} + 4C.
\ee
In terms of the electric an magnetic intensities $E_{i_1\dots i_{2k-1}}$
and $H_{i_1\dots i_{2k-1}}$ and the electric and magnetic inductions
$D_{i_1\dots i_{2k-1}}$ and $B_{i_1\dots i_{2k-1}}$ this implies
\be
E_{i_1\dots i_{2k-1}} B_{i_1\dots i_{2k-1}} = D_{i_1\dots i_{2k-1}}
H_{i_1\dots i_{2k-1}} + {2\over(2k)!}C.
\ee
One may also obtain these results using the Hamiltonian theory described
above. One replaces $\Lambda^2({\Bbb R}^4)$ by $\Lambda^{2k}({\Bbb
R}^{4k})$ and proceeds in an almost identical fashion. The simplest
example of an electric-magnetic duality invariant Lagrangian would then
be the obvious analogue of the Born-Infeld Lagrangian in the form
(\ref{LBI}). This resolves the obvious puzzle of how to generalize the
determinant in (\ref{LBID}) to $2k$-forms for $k>1$.

\sect{The Open SuperString Lagrangian}

In open superstring theory, loop calculations lead to an effective
Lagrangian density which contains a Born-Infeld term
\cite{BerSezPopTow}--\cite{AndTse}~:
\be
{\frak L} = \sqrt{g} \left\{R-2\left(\nabla\phi\right)^2-{1\over
12}e^{2\phi}H^2\right\} + e^{-3\phi} \left\{\sqrt{g}-\sqrt{{\rm
det}\left(g_{\mu\nu} + e^{2\phi}F_{\mu\nu}\right)}\right\},
\ee
where
\be
H = dB + {1\over 4}A\wedge F.
\ee
In four spacetime dimensions, the 3-form $H$ may be eliminated in favour
of the \pagebreak[1] axion $a$ giving
\ba
{\frak L}  & = & \sqrt{g} \left\{R-2\left(\nabla\phi\right)^2 -
2e^{-2\phi}\left(\nabla a\right)^2\right\} + {1\over
4}a\sqrt{g}F_{\mu\nu}\star F^{\mu\nu} \nonumber \\
 & & \label{LSS} \\
 & + & e^{-3\phi} \left\{\sqrt{g}-\sqrt{{\rm det}\left(g_{\mu\nu} +
e^{2\phi}F_{\mu\nu}\right)}\right\}. \nonumber
\ea
Here we are using units where $b=2\pi\alpha^\prime=1$.

First consider the low energy limit. Keeping only terms quadratic in $F$,
the Lagrangian becomes
\ba
L & = & \left\{R-2\left(\nabla\phi\right)^2 - 2e^{-2\phi}\left(\nabla
a\right)^2\right\} + {1\over 4}aF_{\mu\nu}\star F^{\mu\nu} -{1\over
4}e^\phi F_{\mu\nu}F^{\mu\nu} \nonumber \\
 & & \label{LSSQuad} \\
 & = & \left\{R-2\left(\nabla\phi\right)^2 - 2e^{-2\phi}\left(\nabla
a\right)^2\right\} + a{\bf E}\cdot{\bf B} + {1\over 2}e^\phi \left({\bf
E}^2-{\bf B}^2\right). \nonumber
\ea
The electric induction and magnetic intensity are then given by
\be
{\bf D} = e^\phi{\bf E} + a{\bf B} \quad,\quad {\bf H} = e^\phi{\bf B} -
a{\bf E}.
\ee
These constitutive relations may be rewritten as
\be
\left({{\bf E}\atop{\bf H}}\right) =
\underbrace{
\left(
\begin{array}{cc}
e^{-\phi} & -ae^{-\phi} \\
-ae^{-\phi} & e^\phi+a^2e^{-\phi}
\end{array}
\right)}
_{\textstyle\cal M}
\left({{\bf D}\atop{\bf B}}\right).
\label{SL2RConstit}
\ee
Equation (\ref{SL2RConstit}) is invariant under an action of
$SL(2,{\Bbb R})$. To see how this arises, define a complex scalar field
$\lambda$ by
\be
\lambda = a + ie^\phi
\ee
and a complex 2-component vector $\psi$ by
\be
\psi = \left({1\atop -\lambda}\right).
\ee
Then the matrix ${\cal M}$ may be written as
\be
{\cal M} = {\psi\psi^\dagger + c.c.\over\sqrt{{\rm
det}\left(\psi\psi^\dagger + c.c.\right)}}.
\ee
Chosing the first component of $\psi$ to be $1$ fixes the representation
of ${\cal M}$. Equation (\ref{SL2RConstit}) is thus clearly invariant
under the following action of $SL(2,{\Bbb R})$~:
\ba
\psi \rightarrow \psi^\prime \propto \left(S^T\right)^{-1}\psi \quad &
\Rightarrow & \quad {\cal M} \rightarrow \left(S^T\right)^{-1}{\cal
M}S^{-1} \nonumber \\
 & & \\
\left({{\bf D}\atop{\bf B}}\right) \rightarrow S\left({{\bf D}\atop{\bf
B}}\right) \quad & , & \quad \left({{\bf E}\atop{\bf H}}\right)
\rightarrow\left(S^T\right)^{-1}\left({{\bf E}\atop{\bf H}}\right).
\nonumber
\ea
If
\be
S = \left(
\begin{array}{cc}
p & q \\
r & s
\end{array}
\right)
\qquad\mbox{where  }ps-qr=1,
\ee
then the axion and dilaton fields transform under a Mobius transformation
of $\lambda$~:
\be
\lambda \rightarrow {p\lambda+q\over r\lambda+s}.
\label{axdiltransf}
\ee
The $F_{\mu\nu}$ equations of motion are thus guaranteed to be invariant
under this transformation and it is easy to check that the $a$ and $\phi$
equations are also invariant and so is the energy-momentum tensor.

It is interesting to ask whether any of this duality invariance is
preserved when one considers the full Lagrangian (\ref{LSS}), which may
be written as
\ba
L & = & R - 2\left(\nabla\phi\right)^2 - 2e^{-2\phi}\left(\nabla
a\right)^2 + a{\bf E}\cdot{\bf B} \nonumber \\
 & & \\
 & + & e^{-3\phi}\left\{1 - \sqrt{1+e^{4\phi}\left({\bf
B}^2-{\bf E}^2\right)-e^{8\phi}\left({\bf E}\cdot{\bf
B}\right)^2}\right\}. \nonumber
\ea
Hence the electric induction and magnetic intensity are
\ba
{\bf D} & = & {e^\phi{\bf E}+e^{5\phi}\left({\bf E}\cdot{\bf
B}\right){\bf B} \over\sqrt{1+e^{4\phi}\left({\bf B}^2-{\bf
E}^2\right)-e^{8\phi}\left({\bf E}\cdot{\bf B}\right)^2}} + a{\bf B},
\nonumber \\
 & & \label{DHBE2} \\
{\bf H} & = & {e^\phi{\bf B}-e^{5\phi}\left({\bf E}\cdot{\bf
B}\right){\bf E} \over\sqrt{1+e^{4\phi}\left({\bf B}^2-{\bf
E}^2\right)-e^{8\phi}\left({\bf E}\cdot{\bf B}\right)^2}} - a{\bf E}.
\nonumber
\ea
These equations may be solved to give ${\bf E}$ and ${\bf H}$ in terms of
${\bf D}$ \pagebreak[1] and ${\bf B}$~:
\ba
{\bf E} & = & {\left(1+e^{4\phi}{\bf B}^2\right){\bf
D}-\left(a+e^{4\phi}{\bf B}\cdot{\bf D}\right){\bf B} \over e^\phi
\sqrt{\left(1+e^{2\phi}{\bf D}^2\right)\left(1+e^{4\phi}{\bf
B}^2\right)-e^{2\phi}\left(2a{\bf B}\cdot{\bf D}+e^{4\phi}\left({\bf
B}\cdot{\bf D}\right)^2-a^2{\bf B}^2\right)}}, \nonumber \\
 & & \label{EHDB2} \\
{\bf H} & = & {e^{2\phi}\left(1+a^2e^{-2\phi}+e^{2\phi}{\bf
D}^2\right){\bf B}-\left(a+e^{4\phi}{\bf B}\cdot{\bf D}\right){\bf D}
\over e^\phi \sqrt{\left(1+e^{2\phi}{\bf D}^2\right)\left(1+e^{4\phi}{\bf
B}^2\right)-e^{2\phi}\left(2a{\bf B}\cdot{\bf D}+e^{4\phi}\left({\bf
B}\cdot{\bf D}\right)^2-a^2{\bf B}^2\right)}}. \nonumber
\ea
The matrix ${\cal M}$ can then be read off from these equations.
Note that these equations agree with equations (\ref{EHDB}) on setting
$a=\phi=0$. However, for $a\ne 0$, the discrete electric-magnetic duality
invariance that the pure Born-Infeld theory had is no longer apparent.

As was shown above, this theory has an $SL(2,{\Bbb R})$ duality
invariance to lowest order in the electric and magnetic fields. This
determines how all the fields must transform under duality and, if the
full theory has any dualities, they must be a subgroup of the $SL(2,{\Bbb
R})$ duality above. Consider the infinitesimal $SL(2,{\Bbb R})$
transformation described by the matrix
\be
S = \left(
\begin{array}{cc}
1-\alpha & \beta \\
\gamma & 1+\alpha
\end{array}
\right).
\ee
Under this transformation, the fields transform according to~:
\be
\left.
\begin{array}{lcl}
a\rightarrow a(1-2\alpha)+\beta+\gamma(e^{2\phi}-a^2) & , &
e^\phi\rightarrow e^\phi(1-2\alpha-2a\gamma), \\
 & & \\
{\bf E}\rightarrow (1+\alpha){\bf E}-\gamma{\bf H} & , & {\bf
H}\rightarrow (1-\alpha){\bf H}-\beta{\bf E}, \\
 & & \\
{\bf D}\rightarrow (1-\alpha){\bf D}+\beta{\bf B} & , & {\bf
B}\rightarrow (1+\alpha){\bf B}+\gamma{\bf D}.
\end{array}
\right.
\label{Transf}
\ee
It can be shown that these transformations are consistent with the
constitutive relations (\ref{DHBE2}), (\ref{EHDB2}) if and only if
$\alpha=-a\gamma$, that is, in contrast to the previous case, the matrix
$S$ would have to depend on the axion field $a$. However, in that case,
it is easily seen that the axion and dilaton equations of motion,
\be
4\nabla^2a = 8(\nabla a)(\nabla\phi) - e^{2\phi}{\bf E}\cdot{\bf B}
\ee
and
\be
4\nabla^2\phi + 4e^{-2\phi}(\nabla a)^2 = 3e^{-3\phi} -
{3e^{-3\phi}+e^\phi\left({\bf B}^2-{\bf E}^2\right)+e^{5\phi}\left({\bf
E}\cdot{\bf B}\right)^2\over\sqrt{1+e^{4\phi}\left({\bf B}^2-{\bf
E}^2\right)-e^{8\phi}\left({\bf E}\cdot{\bf B}\right)^2}},
\ee
cannot be invariant under these transformations, unless
$\alpha=\gamma=0$.

Another way of seeing that the theory is not duality invariant is to note
that the energy-momentum tensor is not invariant under
these transformations. In particular, it is easy to see that the energy
density $U=T_{00}$ is not invariant. $U$ may be written as a sum of the
contribution from the Born-Infeld term and the contribution from the
kinetic terms for the axion and dilaton, $U=U_{BI}+U_{ax-dil}$. $U_{BI}$
may be calculated by a Legendre transformation of the electromagnetic
part of the Lagrangian~:
\be
U_{BI} = {\bf E}\cdot{\bf D} - L_{em}
\ee
where
\be
L_{em} = a{\bf E}\cdot{\bf B} + e^{-3\phi}\left\{1 -
\sqrt{1+e^{4\phi}\left({\bf B}^2-{\bf E}^2\right)-e^{8\phi}\left({\bf
E}\cdot{\bf B}\right)^2}\right\}.
\ee
The result may be written as
\be
U_{BI} = e^{-3\phi} \left\{\sqrt{1+e^{4\phi}{\bf B}^2+e^{2\phi}\left({\bf
D}-a{\bf B}\right)^2+e^{6\phi}\left({\bf B}\times{\bf D}\right)^2} -
1\right\},
\ee
which is invariant under the transformations (\ref{Transf}) if
$\alpha=-a\gamma$.

However, $U_{ax-dil}$ is not invariant. This can be seen by considering
the matrix defined in equation (\ref{SL2RConstit}),
\be
{\cal M} = \left(
\begin{array}{cc}
e^{-\phi} & -ae^{-\phi} \\
-ae^{-\phi} & e^\phi+a^2e^{-\phi}
\end{array}
\right).
\ee
Note that this matrix transforms under $SL(2,{\Bbb R})$ according to
\be
{\cal M} \rightarrow \left(S^T\right)^{-1} {\cal M} S^{-1}.
\ee
Also note that the kinetic terms for the axion and dilaton may be written
in the form
\be
-2(\nabla\phi)^2-e^{-2\phi}(\nabla a)^2 = 2{\rm det}(\nabla{\cal M}).
\ee
Therefore they are invariant under {\em constant} $SL(2,{\Bbb R})$
transformations and hence so is $U_{ax-dil}$. However, the $SL(2,{\Bbb
R})$ transformation necessary to keep the constitutive relation
invariant (given by $\alpha=-a\gamma$) is not constant and so the kinetic
terms and $U_{ax-dil}$ are not invariant. Hence $U$ is not invariant and
so such an electric-magnetic duality is not possible.

Thus all of the $SL(2,{\Bbb R})$ duality invariance of
the equations of motion of the Lagrangian (\ref{LSSQuad}) is lost when
one considers the full Born-Infeld term of (\ref{LSS}) (except for the
trivial translational invariance of the axion $a$).

\sect{Four Dimensional Spacetime Solutions}

To illustrate our results in 4 dimensions we consider spherically
symmetric gravitating solutions (in the presence of a cosmological term
$\Lambda$) of non-linear electrodynamics. We neglect the axion and
dilaton in this section since no spacetime solutions of the full theory
with the axion and dilaton are yet known. Because the procedure for
obtaining local solutions has been extensively studied in the literature
\cite{Hoff}--\cite{Wil} we merely quote the results we need, which are
easy enough to obtain anyhow. For any such theory, whether dual symmetric
or not, it is known that  Birhoff's theorem holds, and so we may assume
the solution is static and takes the form~:
\be
ds^2 = - \left(1 - {2Gm(r)\over r}\right) dt^2 +
{dr^2 \over 1 - {2Gm(r)\over r}} + r^2 \left(d\theta^2 + \sin^2\theta
d\phi^2\right)
\label{Metric}
\ee
with
\be
D(r) \equiv G_{tr} = {Q\over 4\pi r^2} \quad ,\quad B(r) \equiv
F_{\theta\phi} = {P\over 4\pi} \sin\theta ,
\ee
where $Q$ is the electric and $P$ the magnetic charge and where
\be
{dm \over dr} = {\Lambda\over 2G} r^2 + 4\pi r^2 U\left(D(r),
B(r)\right).
\label{dmdr}
\ee
The function $U({\bf D} , {\bf B})$ is the energy density in the local
orthonormal frame aligned with the static Killing vector. $B(r)$ and
$D(r)$ are the magnetic and electric inductions in that frame. The energy
density may be obtained from the Lagrangian $L= L({\bf E}, {\bf B}) $ by
a Legendre transformation~:
\be
U=-L + {\bf E} \cdot {\bf D},
\ee
where
\be
{\bf D}= {\partial L \over \partial {\bf E}}.
\ee
For a dual-symmetric theory $U$ depends on ${\bf B}$ and ${\bf D}$
only in the dual invariant  combinations ${\bf B}^2 + {\bf D}^2$ and
${\bf B} \times {\bf D}$. Thus, for example, in Born-Infeld theory
\ba
U & = & {1\over b^2} \left\{ \sqrt{1 + b^2 \left({\bf D}^2 + {\bf
B}^2\right) + b^4 \left({\bf D} \times {\bf B}\right)^2} - 1 \right\}
\nonumber \\
 & & \label{UBI} \\
 & = & {1\over b^2} \left\{ \sqrt{1 +
{b^2\left(Q^2+P^2\right)\over\left(4\pi r^2\right)^2}} - 1 \right\}.
\nonumber
\ea
It is clear from (\ref{Metric}), (\ref{dmdr}) and (\ref{UBI}) that in
Born-Infeld theory, the gravitational field of a source with electric
charge $Q$ and magnetic charge $P$, depends only on the dual invariant
combination $Z^2=Q^2+P^2$ and so $g_{\mu\nu}$ is invariant under duality
rotations, which send $Q+iP$ into $e^{i\alpha}(Q+iP)$. This is as expected
since $T_{\mu\nu}$ is unchanged by duality transformations in dual
invariant theories and hence so is the metric. This will not be true if
the theory is not dual invariant. In the case of Born-Infeld theory one
finds that
\be
 - g_{00} = 1 - {2GM\over r} - \Lambda {r^2\over 3} +
{2G\over b^2 r} \int_r^\infty dx \left(\sqrt{(4\pi x^2)^2 + b^2 Z^2} -4\pi
x^2\right) ,
\label{g00}
\ee
where $M$ is a constant of integration. If one wants an asymptotically
flat solution one must set $\Lambda=0$.  In that case $M$ is the ADM
mass. It is helpful to rewrite (\ref{g00}) as
\ba
 - g_{00} = 1 - {2G\over r} \left(M - {1\over b^2} \int^\infty_0 dx
\left(\sqrt{(4\pi x^2)^2 + b^2 Z^2} - 4\pi x^2\right)\right)
\hspace{1.5cm} \nonumber \\
 -\Lambda {r^2\over 3} - {2G\over b^2 r} \int^r_0 dx
\left(\sqrt{(4\pi x^2)^2 + b^2 Z^2} - 4\pi x^2\right).
\label{g00rewrite}
\ea
 From now on we assume that  $\Lambda =0$. Defining $M^\prime$ by
\be
M^\prime  = M - \underbrace{{1\over b^2} \int^\infty_0 dx
\left(\sqrt{(4\pi x^2)^2 + b^2 Z^2} - 4\pi x^2\right)}_{\textstyle\cal E},
\label{Mprime}
\ee
${\cal E}$ has the interpretation of the energy of the electromagnetic
field and so $-M^\prime$ may be interpreted as the binding energy. The
electromagnetic field energy ${\cal E}$ may be re-expressed as
\be
{\cal E} = {\left(Q^2+P^2\right)^{3\over 4}\over\sqrt{4\pi b}}
\int_0^\infty dy\left(\sqrt{y^4+1}-y^2\right).
\label{Energy}
\ee
The $y$ integral may then be integrated by parts becoming a standard
elliptic integral which may be expressed in terms of $\Gamma$ functions~:
\ba
I & = & \int_0^\infty dy\left(\sqrt{1+y^{-4}}-1\right)y^2 \nonumber \\
 & = & {2\over 3} \int_0^\infty {dy\over\sqrt{y^4+1}} \\
 & = & {\pi^{3\over 2}\over 3\Gamma\left({3\over 4}\right)^2} \ = \
1.23604978\dots \nonumber
\ea

The behaviour of the solutions depends on $M^\prime$ and ${2G\over b}
|Z|$. If $M^\prime >0$ there is just one non-degenerate horizon. If
$M^\prime=0$ and ${2G\over b}|Z|\ge 1$ there is also just one
non-degenerate horizon. If $M^\prime=0$ and ${2G\over b}|Z|<1$ there is
no horizon but because  at $r=0$, $g_{rr} \rightarrow 1-{2G\over b}|Z|$,
one has a conical singularity at the origin. This happens because the
energy density is proportional to $ 1\over r^2$ for small $r$. It seems
reasonable to regard the solutions with $M^\prime >0$ as black holes
formed by implosion of the solutions with $M^\prime=0$.

The case  $M^\prime <0$ is more like Reissner-Nordstr\"om  in that there
may be two non-degenerate horizons or one degenerate horizon or no
horizons depending on the size of $M$ relative to $|Z|$ and $b$. Note
that the solutions with one or two horizons always have $M>0$ and
${2G\over b}|Z|>1$.

The fact that the  purely electric solutions with  $M^\prime =0$ and no
horizons have conical singularities was pointed out by Einstein and
Rosen. It  was one of the  reasons why Born abandoned his attempt to
construct a finite classical theory of electrically charged particles
using the Lagrangian (\ref{LBID}). If he had succeeded he knew that  by
dual invariance, he would  also have succeeded in constructing a finite
classical theory of magnetically charged particles . Since he did not
believe that magnetic monopoles exist he chose to abandon (\ref{LBID})
and turn to other Lagrangians for non-linear electrodynamics which are
not dual invariant.

Born's original theory may be viewed in modern terms as an attempt to
find classical solutions representing electrically charged ``elementary
states'' with sources which have finite self-energy. As we remarked above
in a theory admitting duality invariance, even a discrete dual invariance,
there are also  magnetically charged classical solutions representing
elementary states with sources with finite self-energy.

We now close with some speculative remarks. For the Born-Infeld
Lagrangian Born's classical elementary states would have mass
\be
M \approx {1.236e^{3\over 2}\over\sqrt{8\pi^2\alpha^\prime}},
\label{mass formula}
\ee
equating $M$ with ${\cal E}$ given by (\ref{Energy}). Born tried  make
contact with quantum field theory by identifying the particles going
around closed  loops which give rise the Euler-Heisenberg Lagrangian
\cite{EulHei} with his electrically charged elementary states. Equating
the masses gave a consistency condition which turned  out to determine a
value for the fine structure constant. Apart from the fact that this did
not give a numerically accurate value (it gave $\alpha\approx{1\over 82}$)
the procedure was not really consistent because the Euler-Heisenberg
Lagrangian for spinors, or scalars, does not coincide, even at lowest
non-trivial order, with the Born-Infeld Lagrangian. The calculation would
look slightly more convincing if one considered a hypermultiplet
consisting of a charged fermion and two charged bosons going around the
loop because, in this case at least, the Euler-Heisenberg Lagrangian
does coincide with the Born-Infeld Lagrangian at lowest order. Equating
${\cal E}$ given by (\ref{Energy}) with the mass $M$ and using
(\ref{Euler}) gives
\be
\alpha = {e ^2 \over 4 \pi} \approx {1 \over 44},
\label{fine}
\ee
for a purely electrically charged state with electric charge $e$.

In the light of renormalization theory these calculations of the
self-energy of electrically charged particles do not seem very
convincing. Currently there is much speculation that elementary string
states may be related to extreme black holes. Consider a hypermultiplet
of extreme charged black holes with $m={e\over\sqrt{4\pi G}}$. Virtual
black holes going round a closed loop would induce corrections to the
Maxwell Lagrangian. We do not know precisely what form this effective
Lagrangian takes, but it seems reasonable to suppose that, at lowest
order at least, it coincides with the Born-Infeld Lagrangian. This is
because both the bosonic and the superstring Lagrangians have this
property and moreover the behaviour of black holes is classically
invariant under duality rotations and appears to be invariant under
discrete duality transformations at the semi-classical level
\cite{HawRos}.

It is therefore of interest to study the extreme electrically
charged black holes of Born-Infeld theory. These are the solutions with
$M^\prime\le 0$, $M>0$ and ${2G\over b}e>1$. From (\ref{g00rewrite}) the
degenerate horizon of an extreme black hole is located at $r=r_H$, where
$r_H$ satisfies
\be
r_H - 2GM^\prime = f(r_H) \quad ,\quad \left. {df\over dr} \right|
_{r=r_H} = 1
\ee
and
\be
f(r) = {2G\over b^2} \int^r_0 dx \left(\sqrt{(4\pi x^2)^2 + b^2 e^2} -
4\pi x^2\right).
\ee
This implies that
\be
r_H = \sqrt{4G^2 e^2-b^2 \over 16\pi G}.
\ee
Using (\ref{Mprime}) and (\ref{Energy}) then gives the mass $M$ of an
extreme electrically charged Born-Infeld black hole, as a function of its
charge $e$ and of $b$~:
\be
M \approx {1.236e^{3\over 2}\over\sqrt{4\pi b}} + {1\over 2G}\left(r_H -
f(r_H)\right).
\label{Meb}
\ee
For $2Ge\gg b$ this agrees with the charge to mass ratio of the extreme
Reissner-Nordstr\"om black holes of Einstein-Maxwell theory, i.e.\
$M={e\over\sqrt{4\pi G}}$. If $2Ge\gapprox b$ then
$M\approx{1.236e\over\sqrt{8\pi G}}$. The numerical factors
${1\over\sqrt{4\pi}}$ and ${1.236\over\sqrt{8\pi}}$ are approximately the
same and plotting a graph of $M$ defined by (\ref{Meb}) against $e$ and
$b$ shows that the charge to mass ratio of extreme electrically charged
black holes in Born-Infeld theory is always approximately the same as in
Einstein-Maxwell theory.

It thus seems reasonable to assume that the effective Lagrangian obtained
by considering virtual black holes going round closed loops will also
give rise to extreme electrically charged black holes satisfying
$M\approx{e\over\sqrt{4\pi G}}$. Consistency with equation (\ref{Euler})
requires that $b\approx\sqrt{2\over 3}G$. If one then thinks that such
states correspond to open string states, (\ref{alphaprime}) implies that
\be
2 \pi \alpha ^ \prime \approx \sqrt { 2 \over 3} G. \label{ Newton}
\ee
Thus if this approach is justified, the ratio of $\alpha^\prime$ to $G$
in string theory will be constrained.

It is also interesting to note that exact, electrically charged extreme
solutions of Born-Infeld theory have a minimum electric charge given by
$e={b\over 2G}$. For these solutions ${M\over
e}\approx{1.236\over\sqrt{8\pi G}}$ and equation (\ref{Euler}) gives
$\alpha={e^2\over 4\pi}\approx{1\over 44}$ as before. Clearly the precise
details of the calculation above cannot be trusted, however, it raises
the interesting possibility that the coupling constant and the ratio of
$\alpha^\prime$ to $G$ might be determined by consideration of
non-perturbative black hole states. It would therefore be of great
interest to obtain solutions with non-vanishing axion and dilaton fields.

\bigskip
\hfil{\large\bf Acknowledgements}\hfil

\medskip
\noindent
D.A.R was supported by EPSRC grant no.~9400616X.

\end{document}